\documentclass{article}

\usepackage{amsmath}

\usepackage{amssymb}

\begin{document}

\title{The existence of Bogomolny decompositions for gauged baby Skyrme models}

\author{{\L}. T. St\c{e}pie\'{n} \thanks{The Pedagogical University of Cracow, ul. Podchora\c{}\.{z}ych 2, 30-084 Krak\'{o}w, Poland; e-mail: sfstepie@cyf-kr.edu.pl, stepien50@poczta.onet.pl; URL www.ltstepien.up.krakow.pl}}

\date{}

\maketitle

\begin{abstract}
The Bogomolny decompositions (Bogomolny equations) for the gauged baby Skyrme models: restricted and full one, in (2+0)-dimensions, are derived, for some general classes of the potentials.
The conditions, which must be satisfied by the potentials, for each of these mentioned models, are also derived.
\end{abstract}


\hspace{0.65 in}  Bogomolny decomposition, baby Skyrme model

\section{Introduction}
The baby Skyrme model appeared firstly as an analogon (on plane) to the Skyrme model in three-dimensional space. The last one, was introduced by Skyrme in  \cite{Skyrme1961}, \cite{Skyrme1962}, \cite{Skyrme1971}. It is being used for a description of the physics of strong interactions,  in the case of  low-energies \cite{makhankovrybakovsanyuk1993}. The target space of Skyrme model is $SU(2)$, \cite{Skyrme1961}, \cite{Skyrme1962}, \cite{Skyrme1971}, \cite{makhankovrybakovsanyuk1993}, and for baby Skyrme model the target space is $S^{2}$. In these both models, the topological classification of the static field configurations, by their winding numbers, can be done. 
Similarly to the Skyrme model, the following terms appear in the baby Skyrme model: the term of nonlinear $O(3)$ "sigma" model, the quartic term - the analogon of the Skyrme term and the potential. The potential, in baby Skyrme model, must occur, for existence of static  solutions with finite energy. However, the form of the potential is not restricted. Many different forms of the potentials were investigated, for e.g. in \cite{PietteSchroersZakrz1995}, \cite{Karlineretal2008}, \cite{Adametal2009}, \cite{Adametal2010}, \cite{JMSpeight2010}. In \cite{IoannidouLecht2009} noncommutative baby Skyrmions were investigated and in \cite{DomrinLecht2013} exact BPS bound for noncommutative baby Skyrme model has bee obtained.
The problem of peakons and Q-balls in the baby Skyrme model, was studied in \cite{Lis_2011}. 
Bogomolny bound and Bogomolny equations for gauged sigma model were derived in \cite{Schroers}. In \cite{Schroers2} the existence of soliton solutions of Bogomolny kind, in gauged linear sigma model in (2+1) - dimensions, was proved. In \cite{Ghosh}, it was shown that the Bogomolny bound of (1+1)-dimensional gauged sigma models, can be written down by using terms of two conserved charges, similarly to the Bogomolny bound of the BPS dyons in (3+1)-dimensions. Some new Dirac-Born-Infeld extension of BPS Skyrme model was done in \cite{Bednarski}. Gauged version of Faddeev-Skyrme model in (3+1)-dimensions, with Maxwell term, was discussed in \cite{ShnirZhilin}.
In \cite{Schroersetal} some soliton solutions (in the case $V(S^{i}) = 1 - \vec{n} \cdot  \vec{S}, \ (i=1,2,3), \vec{n}=[0,0,1]$) for gauged full baby Skyrme model,  were studied. The lagrangian of the mentioned gauged full baby Skyrme model in (2+1)-dimensions, with some specific form of $V$, is the sum of \cite{Schroersetal},  \cite{Adametal2012}:
"sigma" term $D_{\mu} \vec{S} \cdot D^{\mu} \vec{S}$, Skyrme term $( D^{\mu} \vec{S} \times D^{\nu} \vec{S})^{2}$, usual Maxwell term $F^{2}_{\mu \nu}$ and the potential $V(\vec{S})$, where $\vec{S}$ is three-component vector field, such that $\mid \vec{S} \mid^{2} = 1$, $\lambda > 0$ is a coupling constant, $D_{\mu}\vec{S}=\partial_{\mu}\vec{S} + A_{\mu} (\vec{n} \times \vec{S})$ is the covariant derivative of vector field $\vec{S}$, $F_{\mu\nu}$ is field strength, called also as the curvature and $\vec{n}=[0,0,1]$ is an unit vector and $\mu, \nu = 0, 1, 2$.
The baby Skyrme model has simpler structure, than three-dimensional Skyrme model and so owing to it, we have an opportunity of better studying of the solutions of Skyrme model in (3+1)-dimensions. However, on the other hand, even in the ungauged version of this model, it is still non-integrable, hard, topologically non-trivial and nonlinear field theory. These reasons cause that it is difficult to make analytical studies of this model and so, the investigations of baby Skyrmions are very often numerical. Therefore, the simplification, but of course, keeping us in the class of Skyrme-like models and simultaneously, giving an opportunity for analytical calculations, is important. One may, for example, simplify the problem of solving of field equations, by deriving Bogomolny equations (sometimes called as Bogomol'nyi equations) for these models, mentioned above. 
All solutions of Bogomolny equations are also the solutions of the Euler-Lagrange equations (their order is bigger than the order of Bogomolny equations).  
Bogomolny equations for ungauged restricted baby Skyrme model with the special form of the potential $V=V(S^{3})$ were derived in \cite{Adametal2010}.\\

 In \cite{Stepien2013}, Bogomolny decompositions for both ungauged models: restricted and full baby Skyrme one, were derived. There was also showed that in the case of ungauged restricted baby Skyrme model, Bogomolny decomposition existed for arbitrary potential (in \cite{JMSpeight2010} Bogomolny equations had been obtained for the potential, which was a square of some non-negative function with isolated zeroes, but by another way than used in \cite{Stepien2013}). Next, in \cite{Stepien2013}, it was also showed that for the case of ungauged full baby Skyrme model, the set of the solutions of corresponding Bogomolny equations was some subset of the set of the solutions of Bogomolny equations for ungauged restricted baby Skyrme model.\\
The technique used in \cite{Adametal2012}, for derivation of  Bogomolny equations for gauged restricted baby Skyrme model in (2+0)-dimensions (in the case $V(S^{i}) = 1 - \vec{n} \cdot  \vec{S}, \ (i=1,2,3)$), was firstly applied by Bogomolny in \cite{Bogomolny1976}, among others, for the nonabelian gauge  theory. Independently, the results, similar to some results obtained in \cite{Bogomolny1976}, were obtained in \cite{BelPolSchwTyup1975} and  \cite{Hosoya1978} - in the context of Bogomolny equations, this last paper has been cited only in \cite{Bialyn1979}. This method is based on some proper separation of the terms in the functional of energy. The solutions of Bogomolny equations, found in this way, minimalize the energy functional and saturate Bogomolny bound (Bogomolny bound is an inequality connecting energy functional and topological charge). 

In \cite{Adametal2012}, the Bogomolny equations for the gauged restricted baby Skyrme model, in (2+0)-dimensions, but for the potentials of the form $V(S^{3})$, have been derived and some non-trivial solutions of these equations have been obtained. Independently, in \cite{Stepien2014}, the Bogomolny decomposition for the gauged restricted baby Skyrme model, for the potential $V(S^{3})$ (written down in stereographical variables), obtained by applying so-called concept of strong necessary conditions, has been presented. In \cite{AdamEtal2013} a novel BPS bound for some gauged BPS sumbodel was investigated.

  In this paper we derive Bogomolny equations (we call them as Bogomolny decomposition) for the both gauged baby Skyrme models: restricted and full one, in (2+0)-dimensions, for some general form of the potential. The gauged restricted baby Skyrme model is characterized by absence of $O(3)$-like term in the lagrangian of gauged full baby Skyrme model. We investigate here the case of the more general form of the potentials $V$ (than this one, investigated in  \cite{Adametal2012} and \cite{Stepien2014}), i.e. we look for: Bogomolny decomposition and the condition, which must be satisfied by the potential $V$, in order to existence of the Bogomolny decomposition.

We derive Bogomolny decompositions, for the gauged baby Skyrme models: restricted (this paper contains among others, some generalization of the results presented in \cite{Stepien2014}) and full one, by applying (in contrary to  \cite{Schroersetal}, \cite{Adametal2012} and \cite{AdamEtal2013}) just the concept of strong necessary conditions, firstly presented in \cite{Sokalski1979} and extended in  \cite{Sokalskietal2001}, \cite{Sokalskietal22001}. We derive also the condition, which must be satisfied by the potentials of the form $V$, for which Bogomolny decomposition exists.

The procedure of deriving of Bogomolny decomposition, from the extended concept of strong necessary conditions, has been presented in \cite{Sokalskietal2002}, \cite{Stepien2003} and developed in \cite{Stepienetal2009}.\\
This paper is organized, as follows. In the next subsections of this section, we briefly describe gauged baby Skyrme models:  restricted and full one, and the concept of strong necessary conditions. At the beginning of the section 2, we derive the most general expressions of the density of the topological invariant, needed for our computations. Next, we derive the Bogomolny decompositions for the gauged baby Skyrme models, by using the concept of strong necessary conditions. There are derived also the conditions for the potentials of the gauged baby Skyrme models, which must be satisfied, in the case of Bogomolny decompositions. Section 3 contains a summary.

\subsection{Gauged baby Skyrme models}

In this paper we consider the gauged baby Skyrme models: full and restricted one, with the potential $V=(\vec{S})$. The lagrangian of gauged full baby Skyrme model has the form (in the lagrangian of gauged restricted baby Skyrme model, the $O(3)$-like term is absent):
 
 \begin{equation}
 \mathcal{L} = D_{\mu} \vec{S} \cdot D^{\mu} \vec{S} + \frac{\lambda^{2}}{4}(D_{\mu} \vec{S} \times D_{\nu} \vec{S})^{2} + F^{2}_{\mu \nu} + V,  \label{lagr_full}
 \end{equation}

 where $\vec{S}$ is three-component vector, such that $\mid \vec{S} \mid^{2} = 1$ and $D_{\mu} 
 \vec{S} = \partial_{\mu} \vec{S} + A_{\mu} (\vec{n} \times \vec{S})$ is covariant derivative of vector field $\vec{S}$ and the form of dependance of the potential $V$, on the dependent variables, has not been specified, obviously, it depends on his arguments 
  such that it is a real Lorentzian scalar. \\

 In this paper we consider gauged restricted baby Skyrme model in (2+0) dimensions, with the energy functional of the following form
 
 \begin{equation}
 H = \frac{1}{2} \int d^{2} x \hspace{0.05 in} \mathcal{H} = \frac{1}{2} \int d^{2} x \bigg( \lambda_{0} D_{i} \vec{S} \cdot D^{i} \vec{S} + \frac{\lambda^{2}_{1}}{4} (\epsilon_{kl} D_{k} \vec{S} \times D_{l} \vec{S})^{2} + F^{2}_{kl} + \gamma^{2} V \bigg), \label{energy}
 \end{equation}
 
 where $x_{1}=x, \hspace{0.05 in} x_{2}=y$ and $i, k, l = 1, 2$. We make the stereographic projection
 
 \begin{equation}
 \vec{S} = \bigg[\frac{\omega + \omega^{\ast}}{1+\omega \omega^{\ast}}, \frac{-i \cdot (\omega-\omega^{\ast})}{1+\omega \omega^{\ast}}, \frac{1 - \omega 
 \omega^{\ast}}{1+\omega \omega^{\ast}}\bigg], \ \ i.e. \ \ \omega = \frac{S_{1}+i S_{2}}{1+S_{3}},
 \label{stereograf}
 \end{equation} 
 
 where $\omega=\omega(x,y) \in \mathbb{C}$, $x, y \in \mathbb{R}$ and $\omega(x,y) = u(x,y) + i v(x,y), u, v \in \mathbb{R}$.\\

  The density of energy functional (\ref{energy}), but without $O(3)$ term, has the form after the stereographic projection (this is the hamiltonian of gauged restricted baby Skyrme model) 
 
  \begin{equation}
  \begin{gathered}
   \mathcal{H}=4\lambda_{1} \frac{[i \cdot (\omega_{,x}\omega^{\ast}_{,y}-\omega_{,y}\omega^{\ast}_{,x}) - A_{1} \cdot (\omega_{,y} \omega^{\ast} + \omega
   \omega^{\ast}_{,y}) + A_{2} \cdot (\omega_{,x} \omega^{\ast} + \omega \omega^{\ast}_{,x})]^{2}}{(1+\omega \omega^{\ast})^{4}} + \\
   \lambda_{2}(A_{2,x} - A_{1,y})^{2} + V(\omega, \omega^{\ast}, A_{1}, A_{2}),
  \end{gathered}
  \end{equation}
  
  where after rescalling, the constants $\lambda_{1}, \lambda_{2}$ have been appeared, instead of $\lambda$ and $\gamma$ has been included in $V$ and $\omega_{,x} \equiv \frac{\partial \omega}{\partial x}$, etc. \\

  The Euler-Lagrange equations for this model are, as follows

  \begin{equation}
  \begin{aligned}
  \frac{d}{dx}[N_{1} \cdot (i\omega^{\ast}_{,y}+A_{2}\omega^{\ast})]+\frac{d}{dy}[N_{1} \cdot (-i\omega^{\ast}_{,x}-A_{1}\omega^{\ast})]+
  \frac{1}{4\lambda_{1}}N^{2}_{1}\omega^{\ast}(1+\omega\omega^{\ast})^{3} -\\ 
  N_{1} \cdot (-A_{1}\omega^{\ast}_{,y}+A_{2}\omega^{\ast}_{,x}) - 
  V(\omega,\omega^{\ast},  A_{1}, A_{2})_{,\omega}=0,\\
  c.c.\\
  -2\lambda_{2}\frac{d}{dy}(A_{2,x}-A_{1,y})+N_{1} \cdot (\omega_{,y} \omega^{\ast} + \omega \omega^{\ast}_{,y}) - V(\omega, \omega^{\ast},  A_{1}, A_{2})_{,A_{1}} = 0, \\
  2\lambda_{2}\frac{d}{dx}(A_{2,x}-A_{1,y})-N_{1} \cdot (\omega_{,x} \omega^{\ast} + \omega \omega^{\ast}_{,x}) -  V(\omega, \omega^{\ast},  A_{1}, A_{2})_{,A_{2}} = 0,  
   \end{aligned} 
   \end{equation}
 
  where $N_{1}=\frac{8\lambda_{1}}{(1+\omega\omega^{\ast})^{4}} [i \cdot (\omega_{,x}\omega^{\ast}_{,y}-\omega_{,y}\omega^{\ast}_{,x}) -   
  A_{1}  \cdot (\omega_{,y} \omega^{\ast} + \omega \omega^{\ast}_{,y}) + A_{2}  \cdot (\omega_{,x} \omega^{\ast} + \omega \omega^{\ast}_{,x})]$, \\ 
  $V(\omega, \omega^{\ast},  A_{1}, A_{2})_{,\omega}$ denotes the derivative of $V$ with respect to $\omega$,
  and $\omega_{,x} \equiv \frac{\partial \omega}{\partial x}$, etc.\\  
 
  After making the transformation (\ref{stereograf}), the density of the energy functional (\ref{energy}) has the form (this is the hamiltonian of gauged full baby Skyrme model) 
 
  \begin{equation}
  \begin{gathered}
   \mathcal{H}= \lambda_{00} \frac{(A^{2}_{1}+A^{2}_{2}) \cdot (u^{2}+v^{2}) - 2A_{1}  \cdot (u_{,x}v-u v_{,x}) - 2A_{2}  \cdot (u_{,y}v - uv_{,y}) + u^{2}_{,x} + u^{2}_{,y} + v^{2}_{,x} + v^{2}_{,y}}{(1+u^{2}+v^{2})^{2}} + \\ 
   \lambda_{11} \frac{[(u_{,x}v_{,y}-u_{,y}v_{,x}) - A_{1}  \cdot (u u_{,y} + v v_{,y}) + A_{2}  \cdot (u u_{,x} + v v_{,x})]^{2}}{(1+ u^{2} + v^{2})^{4}} + \\ \label{gaugedFullBabySkyrme1}
   \lambda_{2}(A_{2,x} - A_{1,y})^{2} + V(u,v, A_{1}, A_{2}),
  \end{gathered}
  \end{equation}
  
  where after rescalling, the constants $\lambda_{00}=4\lambda_{0}, \lambda_{11}=16\lambda_{1}, \lambda_{2}$, have been appeared. The constant $\gamma$ has been included in $V$ and $u(x,y)=\Re(\omega(x,y)), v(x,y)=\Im(\omega(x,y)) \in  
  \mathbb{R}$. Of course, in these both gauged baby Skyrme models: restricted and full one, the potentials depend on their arguments such that they are real Lorentzian scalars.\\ 

  The Euler-Lagrange equations for the gauged full baby Skyrme model, have the form

  \begin{gather}
     \frac{d}{dx} \bigg\{2\lambda_{00} \frac{-A_{1}v+u_{,x}}{(1+u^{2}+v^{2})^{2}} +   N_{2} \cdot (v_{,y}+A_{2}u) \bigg\} + \frac{d}{dy} \bigg\{2\lambda_{00} \frac{-A_{2}v+u_{,y}}{(1+u^{2}+v^{2})^{2}} +  N_{2} \cdot (-v_{,x} - A_{1}u) \bigg\} -  \nonumber \\
   2\lambda_{00} \frac{[(A^{2}_{1}+A^{2}_{2})u + A_{1}v_{,x} + A_{2}v_{,y}]}{(1+u^{2}+v^{2})^{2}} -  N_{2} \cdot (-A_{1}u_{,y}+A_{2}u_{,x}) + \frac{2}{\lambda_{11}} u N^{2}_{2} \cdot (1+ u^{2} + v^{2})^{3} +  \nonumber \\  
   4 \lambda_{00} u \frac{(A^{2}_{1}+A^{2}_{2}) \cdot (u^{2}+v^{2}) - 2A_{1} \cdot (u_{,x}v-u v_{,x}) - 2A_{2} \cdot (u_{,y}v - uv_{,y}) + u^{2}_{,x} + u^{2}_{,y} + v^{2}_{,x} +  v^{2}_{,y}}{(1+u^{2}+v^{2})^{3}} - \nonumber \\ 
   V_{,u} = 0,  \\ \label{EL_FullBabySkyrme}
    \ \ {\rm{the \ analogical \  equation, \  following \ from \ varying \ the \ energy \  functional \  with \  respect \  to}}  \ v,  \nonumber \\
    -2\lambda_{2} \frac{d}{dy} (A_{2,x} - A_{1,y})  - 2 \lambda_{00} \frac{A_{1}  \cdot (u^{2}+v^{2}) - (u_{,x} v - u v_{,x})}{(1+u^{2}+v^{2})^{2}} + N_{2} \cdot (uu_{,y}+vv_{,y}) - V_{,A_{1}} = 0,  \nonumber \\
 2\lambda_{2} \frac{d}{dx} (A_{2,x} - A_{1,y}) - 2 \lambda_{00} \frac{A_{2}  \cdot (u^{2}+v^{2}) - (u_{,y} v - u v_{,y})}{(1+u^{2}+v^{2})^{2}} - N_{2} \cdot (uu_{,y}+vv_{,y}) - V_{,A_{2}} = 0, \nonumber
 \end{gather}

  where $N_{2} = \frac{2\lambda_{11}}{(1+u^{2}+v^{2})^{4}}[(u_{,x}v_{,y}-u_{,y}v_{,x}) - A_{1}  \cdot (u u_{,y} + v v_{,y}) + A_{2}  \cdot (u u_{,x} + v v_{,x})]$.\\

 In the case of these both gauged baby Skyrme models: restricted and full one, we assume at the beginning, the dependance of their potentials $V$, on the gauge field $A_{k}, \  (k =1, 2)$, and we want to investigate, whether the conditions for the potentials in these models, in the case of existing of Bogomolny decomposition, will permit such dependance. 
 
 \subsection{The concept of strong necessary conditions}

 The idea of the concept of strong necessary conditions is such that instead of considering of the Euler-Lagrange
 equations, 
 
 \begin{equation}
 F_{,u} - \frac{d}{dx}F_{,u_{,x}} - \frac{d}{dt}F_{,u_{,t}}=0, \label{el}
 \end{equation}
  
 following from the extremum principle, applied to the functional
 
 \begin{equation}
 \Phi[u]=\int_{E^{2}} F(u,u_{,x},u_{,t}) \hspace{0.05 in} dxdt, \label{functional}
 \end{equation}
 
 we consider strong necessary conditions, \cite{Sokalski1979}, \cite{Sokalskietal2001}, \cite{Sokalskietal22001}

 \begin{gather}
   F_{,u}=0, \label{silne1} \\
   F_{,u_{,t}}=0, \label{silne2} \\
   F_{,u_{,x}}=0, \label{silne3}
 \end{gather} 
 
 where $F_{,u} \equiv \frac{\partial F}{\partial u}$, etc.
   
 Obviously, all solutions of the system of the equations (\ref{silne1}) - (\ref{silne3}) satisfy the Euler-Lagrange equation (\ref{el}). However, these solutions, if they exist, are very often trivial. So, in order to avoiding such situation, we make gauge transformation of the functional (\ref{functional})
 
  \begin{equation}
  \Phi \rightarrow \Phi + Inv, \label{gauge_transf}
  \end{equation}
  
 where $Inv$ is such functional that its local variation with respect to $u(x,t)$ vanishes:
 $\delta Inv \equiv 0$.\\

  By virtue of this feature, we have the equivalence of: the Euler-Lagrange equations (\ref{el}) and the Euler-Lagrange equations resulting from  requiring of the extremum of $\Phi + Inv$.
 On the other hand, there is not the invariance of the strong necessary conditions (\ref{silne1}) - (\ref{silne3}), with respect to the gauge transformation (\ref{gauge_transf}) and so, we may expect to obtain non-trivial solutions. As one can noticed, the strong necessary conditions (\ref{silne1}) - (\ref{silne3}) constitute the system of the partial differential equations of the order less than the order of Euler-Lagrange equations (\ref{el}). 


\section{Bogomolny decompositions for gauged baby Skyrme models}

\subsection{Derivation of the general expressions for the density of the topological invariant}
  
  The important step is to construct the general form of the density of the topological invariant for the case of the topology of this model. Some construction of the density of this topological invariant, has been given in \cite{Schroers}, \cite{Yang2001} 
 
 \begin{equation}
 I_{1} = \vec{S} \cdot D_{1} \vec{S} \times D_{2} \vec{S} + F_{12} \cdot (1 - \vec{n} \cdot \vec{S}),
 \end{equation}
 
 where $D_{i}\vec{S}=\partial_{i}\vec{S} + A_{i} \vec{n} \times \vec{S}$, ($i=1,2$), is covariant derivative of vector field $\vec{S}$ and $F_{12}=\partial_{1} A_{2} - \partial_{2} A_{1}$ is magnetic field.
 
 After making the stereographic projection (\ref{stereograf}), we have:
 
  \begin{equation}
 \begin{gathered}
 I_{1} = \frac{1}{(1+\omega\omega^{\ast})^{2}}[2(i \cdot (\omega_{,x}\omega^{\ast}_{,y}-\omega_{,y}\omega^{\ast}_{,x}) -   
  A_{1} \cdot (\omega_{,y} \omega^{\ast} + \omega \omega^{\ast}_{,y}) + A_{2} \cdot (\omega_{,x} \omega^{\ast} + \omega \omega^{\ast}_{,x}))] + \\
  \frac{2 \omega \omega^{\ast}}{1+\omega\omega^{\ast}} (A_{2,x} - A_{1,y}). 
 \end{gathered}
 \end{equation}

 It is useful to generalize the above expression such that there we place some real functions (differentiable at least once) $R_{j}=R_{j}(\omega, \omega^{\ast}, A_{1}, A_{2}), \  (j = 1, 2)$

  \begin{equation}
 \begin{gathered}
 I_{1} =\lambda_{3} \cdot \bigg\{R_{1}(\omega, \omega^{\ast}, A_{1}, A_{2}) \cdot [i \cdot (\omega_{,x}\omega^{\ast}_{,y}-\omega_{,y}\omega^{\ast}_{,x}) 
 - A_{1} \cdot (\omega_{,y} \omega^{\ast} + \omega \omega^{\ast}_{,y}) +\\ 
 A_{2} \cdot (\omega_{,x} \omega^{\ast} + \omega \omega^{\ast}_{,x})] + R_{2}(\omega, \omega^{\ast}, A_{1}, A_{2}) \cdot (A_{2,x} - A_{1,y})\bigg\}. \label{niezmiennik1}
 \end{gathered}
 \end{equation}

  We make the functions $R_{j} \ (j = 1, 2)$, as dependent not only on $\omega, \omega^{\ast}$, but also on $A_{k} \ (k = 1, 2)$, in order to get the most general form of $I_{1}$, as it is possible.
  Next, we look for such conditions for the functions $R_{j} \ (j = 1,2)$, that the expression (\ref{niezmiennik1}) is the density of the topological invariant i.e. its variations with respect to $\omega, \omega^{\ast}, A_{k} \ (k =1, 2)$, always vanish.

  As it turns out, $R_{1} = G'_{1}$ and $R_{2} = G_{1}$, hence, this above expression has the following form
 
 \begin{equation}
 \begin{gathered}
  I_{1} =\lambda_{3} \cdot \bigg\{G'_{1} \cdot [i \cdot (\omega_{,x}\omega^{\ast}_{,y}-\omega_{,y}\omega^{\ast}_{,x}) 
 - A_{1} \cdot (\omega_{,y} \omega^{\ast} + \omega \omega^{\ast}_{,y}) +\\ 
 A_{2} \cdot (\omega_{,x} \omega^{\ast} + \omega \omega^{\ast}_{,x})] + G_{1} \cdot (A_{2,x} - A_{1,y})\bigg\}, \label{niezmiennik2_stere}
 \end{gathered}
 \end{equation}

 where $\lambda_{3}=const$. $G_{1}=G_{1}(\omega \omega^{\ast}) \in \mathbb{R}$ is some arbitrary function differentiable at least twice. $G'_{1}$ denotes the derivative of the function $G_{1}$ with respect to its argument: $\omega\omega^{\ast}$. As we see, here is the generalization 
 in comparison with \cite{Stepien2014}, where $G_{1}=G_{1}(\frac{2 \omega \omega^{\ast}}{1+\omega\omega^{\ast}})$. This generalization makes possible, deriving of Bogomolny decomposition, for more wide class
 of the potentials.\\
 When we need to express (\ref{niezmiennik1}) in real functions $u=\Re{(\omega)}, v=\Im{(\omega)}$, then:

 \begin{equation}
 \begin{gathered}
 I_{1} =\lambda_{3} \cdot \bigg\{G'_{1} \cdot [(u_{,x}v_{,y}-u_{,y}v_{,x}) - A_{1} \cdot (u u_{,y}  + v v_{,y}) +\\ 
 A_{2} \cdot (u u_{,x} + v v_{,x})] + \frac{1}{2} G_{1} \cdot (A_{2,x} - A_{1,y})\bigg\}, \label{niezmiennik2_rz}
 \end{gathered}
 \end{equation}

  where $\lambda_{3}=const$. $G_{1}=G_{1}(u^{2}+v^{2}) \in \mathbb{R}$ is some arbitrary function differentiable at least twice. $G'_{1}$ denotes here the derivative of the function $G_{1}$ with respect to its argument: $u^{2}+v^{2}$.

  When we investigate gauged restricted baby Skyrme model, we will use (\ref{niezmiennik2_stere}), as the form of the density of the topological invariant, and when we investigate gauged full Skyrme model, we will use (\ref{niezmiennik2_rz}). In the next subsections, the symbol "$\cdot$" will be neglected, for a simplicity.

  \subsection{Bogomolny decomposition for gauged restricted baby Skyrme model}

  Now we start to investigate gauged restricted baby Skyrme model. We use stereographical variables $\omega, \omega^{\ast} \in \mathbb{C}$.

  We make the following gauge transformation of $\mathcal{H}$, on the sum of the invariants $\sum_{n=1} I_{n}$ (cf. \cite{Stepien2014})
  
  \begin{equation}
  \begin{gathered}
  \mathcal{H} \longrightarrow \tilde{\mathcal{H}}=4\lambda_{1} \frac{[i(\omega_{,x}\omega^{\ast}_{,y}-\omega_{,y}\omega^{\ast}_{,x}) - 
  A_{1}(\omega_{,y} \omega^{\ast} + \omega \omega^{\ast}_{,y}) + A_{2}(\omega_{,x} \omega^{\ast} + \omega \omega^{\ast}_{,x})]^{2}}{(1+\omega 
  \omega^{\ast})^{4}} + \\
   \lambda_{2}(A_{2,x} - A_{1,y})^{2} + V(\omega, \omega^{\ast}, A_{1}, A_{2}) + \lambda_{3}\bigg\{G'_{1} [(i(\omega_{,x}\omega^{\ast}_{,y}-\omega_{,y}\omega^{\ast}_{,x}) 
 - A_{1}(\omega_{,y} \omega^{\ast} + \omega \omega^{\ast}_{,y}) +\\ 
   A_{2}(\omega_{,x} \omega^{\ast} + \omega \omega^{\ast}_{,x}))] + G_{1} (A_{2,x} - A_{1,y})\bigg\} + D_{x} G_{2} + D_{y} G_{3}, \label{przecech}
  \end{gathered}
  \end{equation}

  where $I_{1}$ is given by (\ref{niezmiennik2_stere}), $I_{2}= D_{x} G_{2}(\omega, \omega^{\ast}, A_{1}, A_{2}), I_{3}=D_{y} G_{3}(\omega, \omega^{\ast}, A_{1}, A_{2}), D_{x} \equiv \frac{d}{dx}, D_{y} \equiv \frac{d}{dy}$. $G_{1}=G_{1}(\omega\omega^{\ast})$
  and $G_{k+1}=G_{k+1}(\omega, \omega^{\ast}, A_{1}, A_{2})$, ($k = 1, 2$), are some functions (differentiable at least twice), which are to be determinated later.
  
   \vspace{0.5 in} 
  
  After applying the concept of strong necessary conditions to (\ref{przecech}), we obtain the so-called dual 
  equations (cf. \cite{Stepien2014})
  
  \begin{eqnarray}
  \begin{gathered}
  \tilde{\mathcal{H}}_{,\omega} : \ 
  -16 \lambda_{1} \frac{[i(\omega_{,x}\omega^{\ast}_{,y}-\omega_{,y}\omega^{\ast}_{,x})-A_{1}(\omega_{,y} \omega^{\ast} + \omega 
  \omega^{\ast}_{,y}) + A_{2}(\omega_{,x} \omega^{\ast} + \omega \omega^{\ast}_{,x})]^{2}\omega^{\ast}}{(1+\omega 
  \omega^{\ast})^{5}} + \\
  \frac{8\lambda_{1}[i(\omega_{,x}\omega^{\ast}_{,y}-\omega_{,y}\omega^{\ast}_{,x})-A_{1}(\omega_{,y} \omega^{\ast} + \omega 
  \omega^{\ast}_{,y}) + A_{2}(\omega_{,x} \omega^{\ast} + \omega 
  \omega^{\ast}_{,x})]}{(1+\omega\omega^{\ast})^{4}}(-A_{1}\omega^{\ast}_{,y}+A_{2}\omega^{\ast}_{,x}) + \\ 
	V_{,\omega} +  \lambda_{3}\bigg\{G''_{1} \omega^{\ast}  
  [i(\omega_{,x}\omega^{\ast}_{,y}-\omega_{,y}\omega^{\ast}_{,x})-A_{1}(\omega_{,y} \omega^{\ast} + \omega 
  \omega^{\ast}_{,y}) + A_{2}(\omega_{,x} \omega^{\ast} + \omega \omega^{\ast}_{,x})] + \label{gorne1}    \\
  G'_{1}(-A_{1}\omega^{\ast}_{,y}+A_{2}\omega^{\ast}_{,x}) + 
  G'_{1} \omega^{\ast} (A_{2,x}-A_{1,y})\bigg\} + D_{x}G_{2,\omega}+D_{y}G_{3,\omega}=0,
   \end{gathered} 
   \end{eqnarray}
  
  \begin{equation}
  \begin{gathered}
  \tilde{\mathcal{H}}_{,\omega^{\ast}}  : \ 
  -16 \lambda_{1} \frac{[i(\omega_{,x}\omega^{\ast}_{,y}-\omega_{,y}\omega^{\ast}_{,x})-A_{1}(\omega_{,y} \omega^{\ast} + \omega 
  \omega^{\ast}_{,y}) + A_{2}(\omega_{,x} \omega^{\ast} + \omega \omega^{\ast}_{,x})]^{2}\omega}{(1+\omega 
  \omega^{\ast})^{5}} + \\
  \frac{8\lambda_{1}[i(\omega_{,x}\omega^{\ast}_{,y}-\omega_{,y}\omega^{\ast}_{,x})-A_{1}(\omega_{,y} \omega^{\ast} + \omega 
  \omega^{\ast}_{,y}) + A_{2}(\omega_{,x} \omega^{\ast} + \omega 
  \omega^{\ast}_{,x})]}{(1+\omega\omega^{\ast})^{4}}(-A_{1}\omega_{,y}+A_{2}\omega_{,x}) + \\ 
         V_{,\omega^{\ast}}  +  \lambda_{3}\bigg\{G''_{1} \omega [i(\omega_{,x}\omega^{\ast}_{,y}-\omega_{,y}\omega^{\ast}_{,x})-A_{1}(\omega_{,y} \omega^{\ast} + \omega 
  \omega^{\ast}_{,y}) + A_{2}(\omega_{,x} \omega^{\ast} + \omega \omega^{\ast}_{,x})] + \label{gorne2}    \\
  G'_{1}(-A_{1}\omega_{,y}+A_{2}\omega_{,x}) + G'_{1} \omega (A_{2,x}-A_{1,y})\bigg\} + D_{x}G_{2,\omega^{\ast}}+D_{y}G_{3,\omega^{\ast}}=0,
  \end{gathered}
  \end{equation}
  
  \begin{equation}
  \begin{gathered}
  \tilde{\mathcal{H}}_{,A_{1}}  : \ 
  \frac{8\lambda_{1}[i(\omega_{,x}\omega^{\ast}_{,y}-\omega_{,y}\omega^{\ast}_{,x})-A_{1}(\omega_{,y} \omega^{\ast} + \omega 
  \omega^{\ast}_{,y}) + A_{2}(\omega_{,x} \omega^{\ast} + \omega 
  \omega^{\ast}_{,x})]}{(1+\omega\omega^{\ast})^{4}}(-\omega_{,y}\omega^{\ast}-\omega\omega^{\ast}_{,y}) +\\ 
	V_{,A_{1}} + \lambda_{3}G'_{1}(-\omega_{,y}\omega^{\ast}-\omega\omega^{\ast}_{,y}) + D_{x}G_{2,A_{1}} + D_{y} G_{3,A_{1}} = 0, 
  \label{dolne5}
  \end{gathered}
  \end{equation}
  
  \begin{equation}
  \begin{gathered}
  \tilde{\mathcal{H}}_{,A_{2}}  : \ 
  \frac{8\lambda_{1}[i(\omega_{,x}\omega^{\ast}_{,y}-\omega_{,y}\omega^{\ast}_{,x})-A_{1}(\omega_{,y} \omega^{\ast} + \omega 
  \omega^{\ast}_{,y}) + A_{2}(\omega_{,x} \omega^{\ast} + \omega 
  \omega^{\ast}_{,x})]}{(1+\omega\omega^{\ast})^{4}}(\omega_{,x}\omega^{\ast}+\omega\omega^{\ast}_{,x}) + \\ 
	V_{,A_{2}} + \lambda_{3}G'_{1}(\omega_{,x}\omega^{\ast}+\omega\omega^{\ast}_{,x}) + D_{x}G_{2,A_{2}} + D_{y} G_{3,A_{2}}  = 0, 
  \label{dolne55}
  \end{gathered}
  \end{equation}
  
  \begin{equation}
  \begin{gathered}
  \tilde{\mathcal{H}}_{,\omega_{,x}}  : \ 
  \frac{8\lambda_{1}[i(\omega_{,x}\omega^{\ast}_{,y}-\omega_{,y}\omega^{\ast}_{,x})-A_{1}(\omega_{,y} \omega^{\ast} + \omega 
  \omega^{\ast}_{,y}) + A_{2}(\omega_{,x} \omega^{\ast} + \omega 
  \omega^{\ast}_{,x})]}{(1+\omega\omega^{\ast})^{4}}(i\omega^{\ast}_{,y}+A_{2}\omega^{\ast}) + \\
  \lambda_{3}G'_{1}(i\omega^{\ast}_{,y}+A_{2}\omega^{\ast}) + G_{2,\omega} = 0, \label{dolne_1} 
  \end{gathered}
  \end{equation}
  
  \begin{equation}
  \begin{gathered}
  \tilde{\mathcal{H}}_{,\omega_{,y}}  : \ 
  \frac{8\lambda_{1}[i(\omega_{,x}\omega^{\ast}_{,y}-\omega_{,y}\omega^{\ast}_{,x})-A_{1}(\omega_{,y} \omega^{\ast} + \omega 
  \omega^{\ast}_{,y}) + A_{2}(\omega_{,x} \omega^{\ast} + \omega 
  \omega^{\ast}_{,x})]}{(1+\omega\omega^{\ast})^{4}}(-i\omega^{\ast}_{,x}-A_{1}\omega^{\ast}) + \\
  \lambda_{3}G'_{1}(-i\omega^{\ast}_{,x}-A_{1}\omega^{\ast}) + G_{3,\omega} = 0, \label{dolne2} 
  \end{gathered}
  \end{equation}
  
  \begin{equation}
  \begin{gathered}
  \tilde{\mathcal{H}}_{,\omega^{\ast}_{,x}}  : \ 
  \frac{8\lambda_{1}[i(\omega_{,x}\omega^{\ast}_{,y}-\omega_{,y}\omega^{\ast}_{,x})-A_{1}(\omega_{,y} \omega^{\ast} + \omega 
  \omega^{\ast}_{,y}) + A_{2}(\omega_{,x} \omega^{\ast} + \omega 
  \omega^{\ast}_{,x})]}{(1+\omega\omega^{\ast})^{4}}(-i\omega_{,y}+A_{2}\omega) + \\
  \lambda_{3}G'_{1}(-i\omega_{,y}+A_{2}\omega) + G_{2,\omega^{\ast}} = 0, \label{dolne3} 
  \end{gathered}
  \end{equation}
  
  \begin{equation}
  \begin{gathered}
  \tilde{\mathcal{H}}_{,\omega^{\ast}_{,y}}  : \ 
  \frac{8\lambda_{1}[i(\omega_{,x}\omega^{\ast}_{,y}-\omega_{,y}\omega^{\ast}_{,x})-A_{1}(\omega_{,y} \omega^{\ast} + \omega 
  \omega^{\ast}_{,y}) + A_{2}(\omega_{,x} \omega^{\ast} + \omega 
  \omega^{\ast}_{,x})]}{(1+\omega\omega^{\ast})^{4}}(i\omega_{,x}-A_{1}\omega) + \\
  \lambda_{3}G'_{1}(i\omega_{,x}-A_{1}\omega) + G_{3,\omega^{\ast}} = 0, \label{dolne4}
  \end{gathered}
  \end{equation}

  \begin{equation}
  \begin{gathered}
  \tilde{\mathcal{H}}_{,A_{1,x}}  : \   G_{2,A_{1}} = 0,
  \label{dolne555}
  \end{gathered}
  \end{equation}
  
  \begin{equation}
  \begin{gathered}
  \tilde{\mathcal{H}}_{,A_{1,y}}  : \  -2\lambda_{2} (A_{2,x} - A_{1,y}) - \lambda_{3} G_{1} + G_{3,A_{1}} = 0,
  \label{dolne5555}
  \end{gathered}
  \end{equation}
  
  \begin{equation}
  \begin{gathered}
  \tilde{\mathcal{H}}_{,A_{2,x}}  : \  2\lambda_{2} (A_{2,x} - A_{1,y}) + \lambda_{3} G_{1} + G_{2,A_{2}} = 0,
  \label{dolne55555}
  \end{gathered}
  \end{equation}

   \begin{equation}
  \begin{gathered}
  \tilde{\mathcal{H}}_{,A_{2,y}}  : \   G_{3,A_{2}} = 0,
  \label{dolne555555}
  \end{gathered}
  \end{equation}
  
  where $G'_{1}, G''_{1}$ denote the derivatives of the function $G_{1}$ with respect to its argument: $\omega\omega^{\ast}$.
  
  \vspace{0.1 in} 
  
  Now, we must make the equations (\ref{gorne1}) - (\ref{dolne555555}) self-consistent.\\
  
  In this order, we need to reduce the number of independent equations by a proper choice of the functions $G_{k}, (k =1, 2, 3)$.
  Very often, such ansatzes exist only for some special $V$ and in most cases of $V$ for many  
  nonlinear field models, it is impossible to reduce the system of corresponding dual equations, to Bogomolny equations. However, even, if we cannot make the reduction mentioned above, 
  such system can be used to derive at least some set of solutions of Euler-Lagrange equations. \\ 
  
  Now, we consider $\omega, \omega^{\ast}, A_{i}, (i=1,2), G_{k}$, ($k=1, 2, 3$), as equivalent dependent variables, governed by the system of equations
  (\ref{gorne1}) - (\ref{dolne555555}). We make two operations (similar operations were made firstly in \cite{Stepien2013}, in the cases of ungauged baby Skyrme models: full and restricted one). 
  
  Namely, as we see, after putting (cf. \cite{Stepien2014})
  
  \begin{gather}
  G'_{1} = -\frac{8\lambda_{1} [i(\omega_{,x}\omega^{\ast}_{,y}-\omega_{,y}\omega^{\ast}_{,x})-A_{1}(\omega_{,y} \omega^{\ast} + \omega 
  \omega^{\ast}_{,y}) + A_{2}(\omega_{,x} \omega^{\ast} + \omega \omega^{\ast}_{,x})]}{\lambda_{3}(1+\omega\omega^{\ast})^{4}}, \label{warG1} \\
  A_{2,x} - A_{1,y} = -\frac{1}{2\lambda_{2}} (\lambda_{3} G_{1}+G_{2,A_{2}}), \label{A2xA1y} \\ 
  G_{3,A_{1}} = - G_{2,A_{2}},  \ \  G_{2}= c_{2}A_{2}, \hspace{0.08 in} G_{3}=-c_{2}A_{1}, \ c_{2}=const, \label{warG2G3}
  \end{gather}

  the equations (\ref{dolne_1}) - (\ref{dolne555555}) become the tautologies and the candidate for Bogomolny decomposition is (cf. \cite{Stepien2014})
  
  \begin{equation}
  \begin{gathered}
  \frac{8\lambda_{1}[i(\omega_{,x}\omega^{\ast}_{,y}-\omega_{,y}\omega^{\ast}_{,x})-A_{1}(\omega_{,y} \omega^{\ast} + \omega 
  \omega^{\ast}_{,y}) + A_{2}(\omega_{,x} \omega^{\ast} + \omega \omega^{\ast}_{,x})]}{\lambda_{3}(1+\omega\omega^{\ast})^{4}} = -G'_{1}, \label{dekBogom_restr}\\
  2\lambda_{2} (A_{2,x} - A_{1,y}) + \lambda_{3} G_{1}+c_{2}= 0.
  \end{gathered}
  \end{equation}
  
  Now, the next step is checking, when the equations (\ref{gorne1}) - (\ref{dolne55}) are satisfied, if (\ref{dekBogom_restr}) hold. Thus, we insert  (\ref{warG2G3}) and (\ref{dekBogom_restr}), 
  into (\ref{gorne1}) - (\ref{dolne55}). From (\ref{dolne5}) - (\ref{dolne55}), we get obtain that $V_{,A_{k}}=0, \ (k=1, 2)$. Hence, we get some system of partial differential 
  equations for $V(\omega, \omega^{\ast})$ and the solution of it is (cf. \cite{Stepien2014})

  \begin{equation}
  \begin{gathered}
  V(\omega, \omega^{\ast}) = \frac{\lambda_{3}}{8\lambda_{1}\lambda_{2}}\bigg(\lambda_{2}\lambda_{3} G'^{2}_{1} (1 + \omega\omega^{\ast})^{4} + \frac{4\lambda_{1}}{\lambda_{3}}(\lambda_{3} G_{1} + c_{2})^{2}\bigg) + \\
 \int \frac{1}{8\lambda_{1}\lambda_{2}}\bigg[ \lambda_{3} \bigg( - \bigg(\int \bigg(\lambda_{3}( 8\lambda_{2} (\omega  \omega^{\ast})^{3} + 18 \lambda_{2} (\omega \omega^{\ast})^{2} + \\ \label{warPoten_restr}
  4(\lambda_{1} + 3\lambda_{2}) \omega  \omega^{\ast} + 2\lambda_{2}) G'^{2} + (\lambda_{2} (9\omega  \omega^{\ast} + 1) (\omega  \omega^{\ast} + 1)^{3} \lambda_{3} G''_{1} +\\
 \omega  \omega^{\ast} \lambda_{2} \lambda_{3} (\omega  \omega^{\ast} + 1)^{4} G'''_{1} + 4\lambda_{1}(\lambda_{3} G_{1} + c_{2})) G'_{1} + G''_{1} \omega^{\ast} (\lambda_{2}\lambda_{3} (\omega  \omega^{\ast} + 1)^{4}G''_{2} + \\
 4\lambda_{1}(\lambda_{3}G_{1} + c_{2}))\omega \bigg) d\omega \bigg) \bigg) \bigg] d \omega^{\ast}.
 \end{gathered}
  \end{equation}


  So, the equations (\ref{dekBogom_restr}) are Bogomolny decomposition for gauged restricted baby Skyrme model in (2+0) dimensions,
  for the potential $V(\omega, \omega^{\ast})$, satisfying (\ref{warPoten_restr}),  where $G_{1}=G_{1}(\omega\omega^{\ast}) \in \mathcal{C}^{3}$ and $G'_{1}, G''_{1}, G'''_{1}$ 
  denote the derivatives of the function $G_{1}$ with respect to its argument: $\omega\omega^{\ast}$.

  \subsection{Bogomolny decomposition for gauged full baby Skyrme model}

   We make gauge transformation of (\ref{gaugedFullBabySkyrme1}), by using two topological invariants of the form (\ref{niezmiennik2_rz}): $\mathcal{H} 
   \rightarrow \tilde{\mathcal{H}}$
  \begin{equation}
  \begin{gathered}
   \tilde{\mathcal{H}}=
   \lambda_{00} \frac{(A^{2}_{1}+A^{2}_{2})(u^{2}+v^{2}) - 2A_{1}(u_{,x}v-u v_{,x}) - 2A_{2}(u_{,y}v - uv_{,y}) + u^{2}_{,x} + u^{2}_{,y} + v^{2}_{,x} + v^{2}_{,y}}{(1+u^{2}+v^{2})^{2}} + \\ 
   \lambda_{11} \frac{[(u_{,x}v_{,y}-u_{,y}v_{,x}) - A_{1}(u u_{,y} + v v_{,y}) + A_{2}(u u_{,x} + v v_{,x})]^{2}}{(1+ u^{2} + v^{2})^{4}} + 
   \lambda_{2}(A_{2,x} - A_{1,y})^{2} + \\ \label{gaugedFullBabySkyrme}
    V(u,v, A_{1}, A_{2}) + \lambda_{3} \{F'_{1}[(u_{,x}v_{,y}-u_{,y}v_{,x}) - A_{1}(u u_{,y} + v v_{,y}) + A_{2}(u u_{,x} + v v_{,x})] +\\ 
   \frac{1}{2}F_{1}(A_{2,x} - A_{1,y})\} + \lambda_{4} \{F'_{2}[(u_{,x}v_{,y}-u_{,y}v_{,x}) - A_{1}(u u_{,y} + v v_{,y}) + A_{2}(u u_{,x} + v v_{,x})] + \\
   \frac{1}{2}F_{2}(A_{2,x} - A_{1,y})\} +   D_{x} G_{3} + D_{y} G_{4},   
  \end{gathered}
  \end{equation}

  where $F_{k}=F_{k}(u^{2}+v^{2}), (k = 1, 2)$ and $G_{n+1}=G_{n+1}(u, v, A_{1}, A_{2}), (n=2, 3)$, are some functions (differentiable at least twice), which are to be determined later and $F'_{k}$ means the derivative of $F_{k}$, with respect to its argument: $(u^{2}+v^{2})$.

  The strong necessary conditions for (\ref{gaugedFullBabySkyrme}), have the form:

   \vspace{-0.2 in} 

  \begin{equation}
  \begin{gathered}
   \tilde{\mathcal{H}}_{,u}  : \ 
   \lambda_{00} \frac{[2(A^{2}_{1}+A^{2}_{2})u + 2A_{1}v_{,x} + 2A_{2}v_{,y}]}{(1+u^{2}+v^{2})^{2}} - \\  
   4 \lambda_{00} u \frac{(A^{2}_{1}+A^{2}_{2})(u^{2}+v^{2}) - 2A_{1}(u_{,x}v-u v_{,x}) - 2A_{2}(u_{,y}v - uv_{,y}) + u^{2}_{,x} + u^{2}_{,y} + v^{2}_{,x} + 
   v^{2}_{,y}}{(1+u^{2}+v^{2})^{3}} + \\ 
  2 \lambda_{11} \frac{[(u_{,x}v_{,y}-u_{,y}v_{,x}) - A_{1}(u u_{,y} + v v_{,y}) + A_{2}(u u_{,x} + v v_{,x})](-A_{1}u_{,y}+A_{2}u_{,x})}{(1+ u^{2} + v^{2})^{4}} -\\
   8 \lambda_{11} u \frac{[(u_{,x}v_{,y}-u_{,y}v_{,x}) - A_{1}(u u_{,y} + v v_{,y}) + A_{2}(u u_{,x} + v v_{,x})]^{2}}{(1+ u^{2} + v^{2})^{5}} + \\ \label{HugaugedFullBabySkyrme}
    V_{,u} + \lambda_{3} \{F'_{1,u}[(u_{,x}v_{,y}-u_{,y}v_{,x}) - A_{1}(u u_{,y} + v v_{,y}) + A_{2}(u u_{,x} + v v_{,x})] +\\ 
   F'_{1}(-A_{1}u_{,y} + A_{2}u_{,x}) + \frac{1}{2} F_{1,u}(A_{2,x} - A_{1,y})\} + \lambda_{4} \{F'_{2,u}[(u_{,x}v_{,y}-u_{,y}v_{,x}) - A_{1}(u u_{,y} + v v_{,y}) + \\ 
   A_{2}(u u_{,x} + v v_{,x})] +  F'_{2}(-A_{1}u_{,y} + A_{2}u_{,x}) + \frac{1}{2} F_{2,u}(A_{2,x} - A_{1,y})\} +  D_{x} G_{3, u} + D_{y} G_{4, u} = 0,
 \end{gathered}
 \end{equation}

\begin{equation}
  \begin{gathered}
\tilde{\mathcal{H}}_{,v}  : \ 
   \lambda_{00} \frac{[2(A^{2}_{1}+A^{2}_{2})v - 2A_{1}u_{,x} -  2A_{2}u_{,y}]}{(1+u^{2}+v^{2})^{2}} - \\  
 4 \lambda_{00} v \frac{[(A^{2}_{1}+A^{2}_{2})(u^{2}+v^{2}) - 2A_{1}(u_{,x}v-u v_{,x}) - 2A_{2}(u_{,y}v - uv_{,y}) + u^{2}_{,x} + u^{2}_{,y} + v^{2}_{,x} + v^{2}_{,y}]}{(1+u^{2}+v^{2})^{3}} + \\ 
  2 \lambda_{11} \frac{[(u_{,x}v_{,y}-u_{,y}v_{,x}) - A_{1}(u u_{,y} + v v_{,y}) + A_{2}(u u_{,x} + v v_{,x})](-A_{1}v_{,y}+A_{2}v_{,x})}{(1+ u^{2} + v^{2})^{4}} -\\
   8 \lambda_{11} v \frac{[(u_{,x}v_{,y}-u_{,y}v_{,x}) - A_{1}(u u_{,y} + v v_{,y}) + A_{2}(u u_{,x} + v v_{,x})]^{2}}{(1+ u^{2} + v^{2})^{5}} + \\ \label{HvgaugedFullBabySkyrme}
    V_{,v} + \lambda_{3} \{F'_{1,v}[(u_{,x}v_{,y}-u_{,y}v_{,x}) - A_{1}(u u_{,y} + v v_{,y}) + A_{2}(u u_{,x} + v v_{,x})] +\\ 
F'_{1}(-A_{1}v_{,y} + A_{2}v_{,x}) + \frac{1}{2} F_{1,v}(A_{2,x} - A_{1,y})\} + \lambda_{4} \{F'_{2,v}[(u_{,x}v_{,y}-u_{,y}v_{,x}) - A_{1}(u u_{,y} + v v_{,y}) +\\ 
A_{2}(u u_{,x} + v v_{,x})] + F'_{2}(-A_{1}v_{,y} + A_{2}v_{,x}) + \frac{1}{2} F_{2,v}(A_{2,x} - A_{1,y})\} +  D_{x} G_{3, v} + D_{y} G_{4, v} = 0, 
 \end{gathered}
  \end{equation}

 \begin{equation}
  \begin{gathered}
\tilde{\mathcal{H}}_{,A_{1}}  : \  \lambda_{00} \frac{2A_{1}(u^{2}+v^{2}) - 2(u_{,x} v - u v_{,x})}{(1+u^{2}+v^{2})^{2}} - \\
2\lambda_{11}\frac{[(u_{,x}v_{,y}-u_{,y}v_{,x}) - A_{1}(u u_{,y} + v v_{,y}) + A_{2}(u u_{,x} + v v_{,x})](uu_{,y}+vv_{,y})}{(1+u^{2}+v^{2})^{4}} + V_{,A_{1}} -\\  \label{HA1gaugedFullBabySkyrme}
\lambda_{3} F'_{1}(uu_{,y}+vv_{,y}) -  \lambda_{4} F'_{2}(uu_{,y}+vv_{,y}) + D_{x}G_{3,A_{1}} + D_{y}G_{4,A_{1}} = 0, \\
 \end{gathered}
  \end{equation}

 \begin{equation}
  \begin{gathered}
\tilde{\mathcal{H}}_{,A_{2}}  : \  \lambda_{00} \frac{2A_{2}(u^{2}+v^{2}) - 2(u_{,y} v - u v_{,y})}{(1+u^{2}+v^{2})^{2}} + \\
2\lambda_{11}\frac{[(u_{,x}v_{,y}-u_{,y}v_{,x}) - A_{1}(u u_{,y} + v v_{,y}) + A_{2}(u u_{,x} + v v_{,x})](uu_{,x}+vv_{,x})}{(1+u^{2}+v^{2})^{4}} + V_{,A_{2}} + \\  \label{HA2gaugedFullBabySkyrme}
\lambda_{3} F'_{1}(uu_{,x}+vv_{,x}) +  \lambda_{4} F'_{2}(uu_{,x}+vv_{,x}) + D_{x}G_{3,A_{2}} + D_{y}G_{4,A_{2}} = 0, 
\end{gathered}
  \end{equation}

 \begin{equation}
  \begin{gathered}
\tilde{\mathcal{H}}_{,u_{x}}  : \  2\lambda_{00} \frac{-A_{1}v+u_{,x}}{(1+u^{2}+v^{2})^{2}} +\\ 
2\lambda_{11} \frac{[(u_{,x}v_{,y}-u_{,y}v_{,x}) - A_{1}(u u_{,y} + v v_{,y}) + A_{2}(u u_{,x} + v v_{,x})](v_{,y}+A_{2}u)}{(1+u^{2}+v^{2})^{4}} + \label{dolne1} \\
\lambda_{3} \{F'_{1} [v_{,y} + A_{2} u] \} + \lambda_{4} \{F'_{2} [v_{,y} + A_{2} u] \} + G_{3,u} = 0, 
\end{gathered}
  \end{equation}

\begin{equation}
  \begin{gathered}
\tilde{\mathcal{H}}_{,u_{y}}  : \  2\lambda_{00} \frac{-A_{2}v+u_{,y}}{(1+u^{2}+v^{2})^{2}} + \\
2\lambda_{11} \frac{[(u_{,x}v_{,y}-u_{,y}v_{,x}) - A_{1}(u u_{,y} + v v_{,y}) + A_{2}(u u_{,x} + v v_{,x})](-v_{,x} - A_{1}u)}{(1+u^{2}+v^{2})^{4}} + \label{dolne2} \\
\lambda_{3} \{F'_{1} [-v_{,x} - A_{1} u] \} + \lambda_{4} \{F'_{2} [-v_{,x} - A_{1} u] \} + G_{4,u} = 0,
\end{gathered}
  \end{equation}

  \begin{equation}
  \begin{gathered}
\tilde{\mathcal{H}}_{,v_{x}}  : \  2\lambda_{00} \frac{A_{1}u+v_{,x}}{(1+u^{2}+v^{2})^{2}} + \\
2\lambda_{11} \frac{[(u_{,x}v_{,y}-u_{,y}v_{,x}) - A_{1}(u u_{,y} + v v_{,y}) + A_{2}(u u_{,x} + v v_{,x})](-u_{,y} + A_{2}v)}{(1+u^{2}+v^{2})^{4}} + \label{dolne3} \\
\lambda_{3} \{F'_{1} [-u_{,y} + A_{2} v] \} + \lambda_{4} \{F'_{2} [-u_{,y} + A_{2} v] \} + G_{3,v} = 0,
  \end{gathered}
  \end{equation}

   \begin{equation}
  \begin{gathered}
\tilde{\mathcal{H}}_{,v_{y}}  : \  2\lambda_{00} \frac{A_{2}u+v_{,y}}{(1+u^{2}+v^{2})^{2}} + \\
2\lambda_{11} \frac{[(u_{,x}v_{,y}-u_{,y}v_{,x}) - A_{1}(u u_{,y} + v v_{,y}) + A_{2}(u u_{,x} + v v_{,x})](u_{,x} - A_{1}v)}{(1+u^{2}+v^{2})^{4}} + \label{dolne4} \\
\lambda_{3} \{F'_{1} [u_{,x} - A_{1} v] \} + \lambda_{4} \{F'_{2} [u_{,x} - A_{1} v] \} + G_{4,v} = 0,
  \end{gathered}
  \end{equation}

  \begin{gather}
   \tilde{\mathcal{H}}_{,A_{1,x}}  : \  G_{3, A_{1}} = 0, \label{dolnef5} \\
   \tilde{\mathcal{H}}_{,A_{1,y}}  : \  -2\lambda_{2} (A_{2,x} - A_{1,y}) - \frac{\lambda_{3}}{2} F_{1} -  \frac{\lambda_{4}}{2} F_{2} + G_{4, A_{1}} = 0, \label{dolne6}\\
   \tilde{\mathcal{H}}_{,A_{2,x}}  : \  2\lambda_{2} (A_{2,x} - A_{1,y}) + \frac{\lambda_{3}}{2} F_{1} +  \frac{\lambda_{4}}{2} F_{2} + G_{3, A_{2}} = 0, \label{dolne7} \\
   \tilde{\mathcal{H}}_{,A_{2,y}}  : \  G_{4, A_{2}} = 0. \label{dolne8}
   \end{gather}

  Now we need to make the equations (\ref{HugaugedFullBabySkyrme}) - (\ref{dolne8}) self-consistent. In this order, at first we put
  \begin{gather}
  u_{,x} + v_{,y} = -\frac{(1+u^{2}+v^{2})^{2}}{2\lambda_{00}} G_{3,u} + A_{1} v - A_{2} u, \label{kand1}\\
  u_{,y} - v_{,x} = \frac{(1+u^{2}+v^{2})^{2}}{2\lambda_{00}} G_{3,v} + A_{1} u + A_{2} v,\\
   u_{,x}v_{,y} - u_{,y}v_{,x} - A_{1} (uu_{,y} + vv_{,y}) + A_{2} (uu_{,x} + vv_{,x}) = -\frac{\lambda_{4}}{2\lambda_{11}} (1+u^{2}+v^{2})^{4} F'_{2},\\
   A_{2,x} - A_{1,y} = -\frac{1}{2\lambda_{2}} \bigg(\frac{\lambda_{3}}{2} F_{1} + \frac{\lambda_{4}}{2} F_{2} + G_{3,A_{2}}\bigg),\\
   F'_{1} = \frac{2\lambda_{00}}{\lambda_{3} (1 + u^{2} + v^{2})^{2}}, \ \  G_{3,uA_{1}}=0, \ \ G_{4,uA_{2}}=0,  \label{kand5}
  \end{gather} 

   where $F'_{1}$ denotes the derivative of the function $F_{1}$, with respect to its argument: $1+u^{2}+v^{2}$.

   Then it has turned out that 

   \begin{equation}
   G_{3,u} = G_{4,v}, \ \ G_{3,v} = - G_{4,u}, \ \  G_{4,A_{1}} = - G_{3,A_{2}}. \label{f1}  
   \end{equation}   
  
    Hence, from  (\ref{dolnef5}) and (\ref{dolne8})

   \begin{equation}
   G_{3} = f(u,v) + c_{2}A_{2}, \  G_{4} = f(u,v) - c_{2}A_{1}, \ \  c_{2} = const, \ \  f_{,uu} + f_{,vv} = 0. \label{f2}
   \end{equation} 
 
  Hence, the equations (\ref{dolne1}) - (\ref{dolne8}) become the tautologies. \\
  The equations (\ref{HA1gaugedFullBabySkyrme}) - (\ref{HA2gaugedFullBabySkyrme}), after taking into account (\ref{dolnef5}) - (\ref{dolne8}), (\ref{kand1}) - (\ref{kand5}), (\ref{f2}) 
  and the fact that the potential $V$ should be a Lorentzian scalar, implicate that $V_{,A_{k}}=0, \ (k=1, 2)$. 
  Hence, after eliminating all expressions including the derivatives of the fields $u, v, A_{1}, A_{2}$, from the equations (\ref{HugaugedFullBabySkyrme}) - (\ref{HA2gaugedFullBabySkyrme}), 
  by using (\ref{kand1}) - (\ref{kand5}) (after taking into account (\ref{f2})), we obtain the system of the partial differential equations for $V(u,v)$ and $f(u,v)$. The solutions 
  of this system are: $f(u,v) = const.$ and the condition for the potential

\begin{gather}
  V(u,v) = \int \frac{1}{\lambda_{11} \lambda_{2} (1+ u^{2} +v^{2})^{3}} ((2 \lambda_{2} \lambda^{2}_{4} (1+u^{2} + v^{2})^{6}
 F'^{2}_{2} +  \nonumber \\
 \lambda_{4} (\lambda_{2}\lambda_{4} (1+u^{2} + v^{2})^{5} F''_{2} +
 \frac{1}{2}(\frac{1}{2} \lambda_{4}(1+u^{2} + v^{2})F_{2} +  \nonumber \\ 
c_{2}(1+u^{2} + v^{2}) -\lambda_{00})\lambda_{11})
(1+u^{2} + v^{2})^{2} F'_{2} +
 (\frac{1}{2}\lambda_{4}(1+u^{2} + v^{2})F_{2} +  \nonumber \\
 c_{2} (1+u^{2} + v^{2}) - \lambda_{00})\lambda_{00}\lambda_{11})u)du + \nonumber \\
 \int \frac{1}{\lambda_{11}\lambda_{2} (1+u^{2} + v^{2})^{3}} \bigg\{ \bigg [ -\frac{1}{2} (1+u^{2} + v^{2})^{3}
\bigg (\int \frac{1}{(1+u^{2} + v^{2})^{4}} \nonumber \\ 
(4(6\lambda^{2}_{4}(\lambda_{2} u^{4} + 2\lambda_{2}(1 + v^{2}) u^{2} + 2 \lambda_{2} v^{2} + \frac{1}{24} \lambda_{11} + (1 + v^{4})\lambda_{2} )(1+u^{2} + v^{2})^{4} F'^{2}_{2} + \nonumber \\ 
\lambda_{4}(8\lambda_{2}\lambda_{4}(1+u^{2} + v^{2})^{5} F''_{2} + \lambda_{2}\lambda_{4} (1+u^{2} + v^{2})^{6} F'''_{2} + \lambda_{00}\lambda_{11}) \nonumber \\
 (1+u^{2} + v^{2})^{2} F'_{2} + \lambda_{2} \lambda^{2}_{4} (1+u^{2} + v^{2})^{8} F''^{2}_{2} + \label{warPoten_full} \\ 
\frac{1}{2}\lambda_{11}\lambda_{4}(1+u^{2} + v^{2})^{3}(\frac{\lambda_{4}}{2}(1+u^{2} + v^{2})F_{2}+c_{2}(1+u^{2} + v^{2}) - \lambda_{00})F''_{2} - \nonumber \\
2\lambda_{00} \lambda_{11} (\frac{\lambda_{4}}{2}(1+u^{2} + v^{2}) F_{2} +
c_{2}(1+u^{2} + v^{2}) - \frac{3}{2}\lambda_{00}))u)du \bigg) +  \nonumber \\ 
2 \lambda_{2}\lambda^{2}_{4}(1+u^{2} + v^{2})^{6} F'^{2}_{2} + \lambda_{4} (\lambda_{2}\lambda_{4} (1+u^{2} + v^{2})^{5} F''_{2} +  \nonumber \\ 
\frac{\lambda_{11}}{2} (\frac{\lambda_{4}}{2} (1+u^{2} + v^{2}) F_{2} + c_{2}(1+u^{2} + v^{2}) - \lambda_{00})) (1+u^{2} + v^{2})^{2} F'_{2} + \nonumber \\ 
\lambda_{00}\lambda_{11}(\frac{\lambda_{4}}{2}(1+u^{2} + v^{2}) F_{2} + c_{2}(1+u^{2} + v^{2}) - \lambda_{00})\bigg ] v \bigg\} dv + C_{1},  \nonumber 
\end{gather}
  
 where $C_{1}=const, F_{2} = F_{2}(u^{2} + v^{2}) \in \mathcal{C}^{3}$ and $F'_{2}, F''_{2}, F'''_{2}$ denote the derivatives of the function $F_{2}$, with respect to its argument: $u^{2}+v^{2}$.

 Hence, the Bogomolny decomposition for gauged full baby Skyrme model in (2+0)-dimensions, has the form:

  \begin{equation}
  \begin{gathered}
  u_{,x} + v_{,y} =  A_{1} v - A_{2} u,\\
  u_{,y} - v_{,x} =   A_{1} u + A_{2} v,\\
  u_{,x}v_{,y} - u_{,y}v_{,x} - A_{1}  (uu_{,y} + vv_{,y}) + A_{2}  (uu_{,x} + vv_{,x}) = -\frac{\lambda_{4}}{2\lambda_{11}} (1+u^{2}+v^{2})^{4} F'_{2}, \label{dekBogom_full}\\
 A_{2,x} - A_{1,y} = -\frac{1}{2\lambda_{2}} \bigg(\frac{\lambda_{4}}{2} F_{2} - \frac{\lambda_{00}}{1+u^{2}+v^{2}} + c_{2}\bigg),
 \end{gathered}
 \end{equation}

  where $F_{2}=F_{2}(u^{2}+v^{2})$ and $F'_{2}$ denotes the derivative of the function $F_{2}$, with respect to its argument: $u^{2}+v^{2}$, and the potential $V(u, v)$ needs to 
  satisfy the conditon (\ref{warPoten_full}).

 \section{Summary}
 
  We started from finding the most general form of the functions\\ 
 $R_{j} = R_{j}(\omega, \omega^{\ast}, A_{1}, A_{2}), \ (j=1, 2)$, in the density of the topological invariant (\ref{niezmiennik1}), written down in complex field variables, and  the most general form of these functions in the density of the topological invariant (an analogon to  (\ref{niezmiennik1})), written down in real field variables $u, v$. It has turned out that $R_{1} = G'_{1}$ and $R_{2} = G_{1}$, where $G_{1}=G_{1}(\omega \omega^{\ast})$ (or $G_{1} = G_{1}(u^{2} + v^{2})$, then the factor $1/2$ appears, by the function $G_{1}(u^{2}+v^{2})$, in the density of the corresponding topological invariant). The form of the dependance of the function $G_{1}$, on the field variables $\omega, \omega^{\ast}$ (or $u, v$) and the independance of $G_{1}$ on $A_{k} \ (k=1, 2)$, have the influence on the dependance of the potential $V$ on these field variables. 
  Next, we applied the concept of strong necessary conditions for gauged restricted baby Skyrme model in (2+0)-dimensions. In result, we obtained Bogomolny decomposition (\ref{dekBogom_restr}), i.e. Bogomolny equations, for this model, for more wide class of the potentials: $V(\omega\omega^{\ast})$, than Bogomolny equations, obtained in \cite{Adametal2012} and  \cite{Stepien2014}. We derived also the condition (for the potential) of existence of this Bogomolny decomposition, this condition has the form (\ref{warPoten_restr}).
We obtained analogical results for the gauged full baby Skyrme model in (2+0)-dimensions, in this case the Bogomolny decomposition has the form (\ref{dekBogom_full}) and the potential needs to satisfy the condition (\ref{warPoten_full}). We see that analogically to \cite{Stepien2013}, where Bogomolny decomposition for ungauged baby Skyrme models: restricted and full one, have been derived, the set of the solutions of Bogomolny decomposition of gauged full baby Skyrme model is the subset of the solutions of Bogomolny decomposition of gauged restricted baby Skyrme model. Moreover, at the beginning of this paper, we have assumed that for the gauged full baby Skyrme model and for the gauged restricted baby Skyrme model, the potentials in their hamiltonians, depend on $\omega, \omega^{\ast}, A_{1}, A_{2}$ and $u, v, A_{1}, A_{2}$, respetively. Next, it has turned out that the most general forms of the topological invariant for these models, are built on, among others, function $G_{1}$ and its derivative with respect to the argument of $G_{1}$: $\omega \omega^{\ast}$ and $u^{2} + v^{2}$, respectively. Finally, this function and its derivatives, have been included into the expression, which $V$ needs to be equal to, if we want to get Bogomolny decomposition. On the other hand, it has turned out that in the case of existence of the Bogomolny decompositions for any of the gauged baby Skyrme model: restricted and full one, the potential $V$ does not depend on $A_{k}, k=1,2$. Hence, in the case of the Bogomolny decompositions for these both models, the potentials of these models cannot include the expression $A_{k}A^{k} \ (k=1, 2)$, which occurs in the potential in  Proca theory \cite{ArodzHadasz2010} or in the theory of a massive vector field \cite{Rubakov2002}.
  \section{Acknoweledgements}
  
  The author thanks to Dr. Hab. A. Wereszczy\'{n}ski for interesting discussions about gauged restricted baby Skyrme model, carried out in 2010. The author thanks also to Dr. Z. Lisowski for some interesting remarks.
  
  \section{Computational resources}
  
    The computations were carried out by using WATERLOO MAPLE Software on computer ``mars'' in ACK-CYFRONET AGH in Krak\'{o}w.
    This research was supported in part by PL-Grid Infrastructure, too.

     \bibliography{L_T_Stepien_BgmGaugbabySkyr}

    \bibliographystyle{unsrt}

\end{document}